\documentclass[prl,onecolumn,superscriptaddress,nofootinbib]{revtex4-1}
\usepackage{graphicx}
\usepackage{color}
\usepackage{amsmath}
\usepackage{amsfonts}
\usepackage{mathrsfs}

\begin{document}

\title[Phyllotaxis on curved surfaces]{Phyllotaxis on surfaces of constant Gaussian curvature}

\author{Jean-Fran\c cois Sadoc}
\email{sadoc@lps.u-psud.fr}
\affiliation{Laboratoire de Physique des Solides (CNRS-UMR 8502), B{\^a}t. 510, Universit{\'e} Paris-sud, F 91405 Orsay cedex}

\author{Jean Charvolin}
\affiliation{Laboratoire de Physique des Solides (CNRS-UMR 8502), B{\^a}t. 510, Universit{\'e} Paris-sud, F 91405 Orsay cedex}

\author{ Nicolas Rivier }
\affiliation{ IPCMS, Universit\'e Louis Pasteur,  F-67084 Strasbourg, cedex }

\begin{abstract}
A close packed organization with circular symmetry of a large number of small discs on a plane is obtained when the centres of the discs are distributed according to the algorithm of phyllotaxis. We study here the distributions obtained on surfaces of constant Gaussian curvatures, positive for the sphere or negative for the hyperbolic plane. We examine how the properties of homogeneity, isotropy and self-similarity typical of the distribution on the plane, and resulting from the presence of circular grain boundaries with quasicrystalline sequences, are affected by the curvature of the bearing surface. The quasicrystalline sequences of the grain boundaries appear indeed to be structural invariants, but the widths of the grains they separate vary differently with the curvature of the surface. The self similarity of the whole organization observed on the plane is therefore lost on the hyperbolic plane and the sphere. The evolutions of the local order within the grains show no differences except on the equatorial belt of the sphere where the isotropy is decreased owing to the symmetry of this finite surface around its equator.
\end{abstract}

\maketitle

\section{Introduction }

The densest organization of small discs on an infinite plane is obtained when their centres are at the nodes of the triangular tiling of the hexagonal lattice. All these discs occupy the same area on the plane and have the same local environment of six equally-spaced first neighbours which is reproduced all over the plane according to the laws of classical crystallography. This crystalline solution remains valid for a limited assembly of small discs within a finite compact domain of the plane under the condition that the borders of the domain be aligned along reticular directions of the lattice. If not, as for a circular domain, this solution is no longer valid, the circular symmetry of the domain being not accounted for by the finite number of rotational symmetries of a crystal.

It has been shown that the best packing efficiency found in the case of circular symmetry is that obtained when the centres of the discs are regularly placed on the spiral drawn by the algorithm of phyllotaxis~\cite{jean1,jean2,ridley}. This solution  approaches circular symmetry, but the area per disc, the number of first neighbours and their distances are no longer constant as they are in the case with hexagonal symmetry. Those differences between crystalline and phyllotactic organizations were recently analyzed considering not only the positions of the centres of the discs but also the polygonal nature of their Voronoi cells~\cite{sadocriviercharvolin}. Such an analysis shows that the phyllotactic best packing efficiency results from an interplay between metric and topological distortions leading to a  structure with an inflation-deflation symmetry.
We study here the distributions of points with their Voronoi cells drawn by the algorithm of phyllotaxis onto surfaces of constant Gaussian curvatures, positive for the sphere or negative for the hyperbolic plane. We examine how the best packing efficiency typical of the distribution on the plane evolves with the curvature of the underlying surface, considering both short-range (the shape of a cell, its area and the distances from its centre to the first neighbour points) and long-range characteristics (the repetition law governing the organization of the cells).

The part of this work concerning plane and spherical phyllotaxis was motivated by our interest for the lateral organizations of long biological molecules in dense fibres with a circular section imposed by their surface tension \cite{charvolinsadoc}. These organizations can indeed be represented by distributions of points on planar or spherical bases according to the fact that the molecules can be parallel or twisted within the fibres. Spherical phyllotaxis was also used earlier as a simple ideal approached by vegetal forms ~\cite{dixon}, in order to obtain the most homogeneous sampling of points on a sphere in numerical integration~\cite{hannay}, to develop climate models of the earth~\cite{swinbank}  or to estimate the Earth coverage of satellite constellations~\cite{gonzalez}. We do not know of any application concerning phyllotaxis on surfaces with negative Gaussian curvature up to now.

Since the distributions on the sphere and the hyperbolic plane will be compared to that observed on the plane, we recall first the main features of the latter, limiting ourselves to what is needed for this comparison, with demonstrations and more details given in~\cite{sadocriviercharvolin}.

\section{Phyllotaxis on  the Euclidean plane}

The Voronoi cells, used in~\cite{sadocriviercharvolin} to describe  phyllotaxis on the plane are an efficient tool to analyze the geometry and the topology of the structure.

\subsection{Pattern}

\begin{figure}[tbp]
\includegraphics{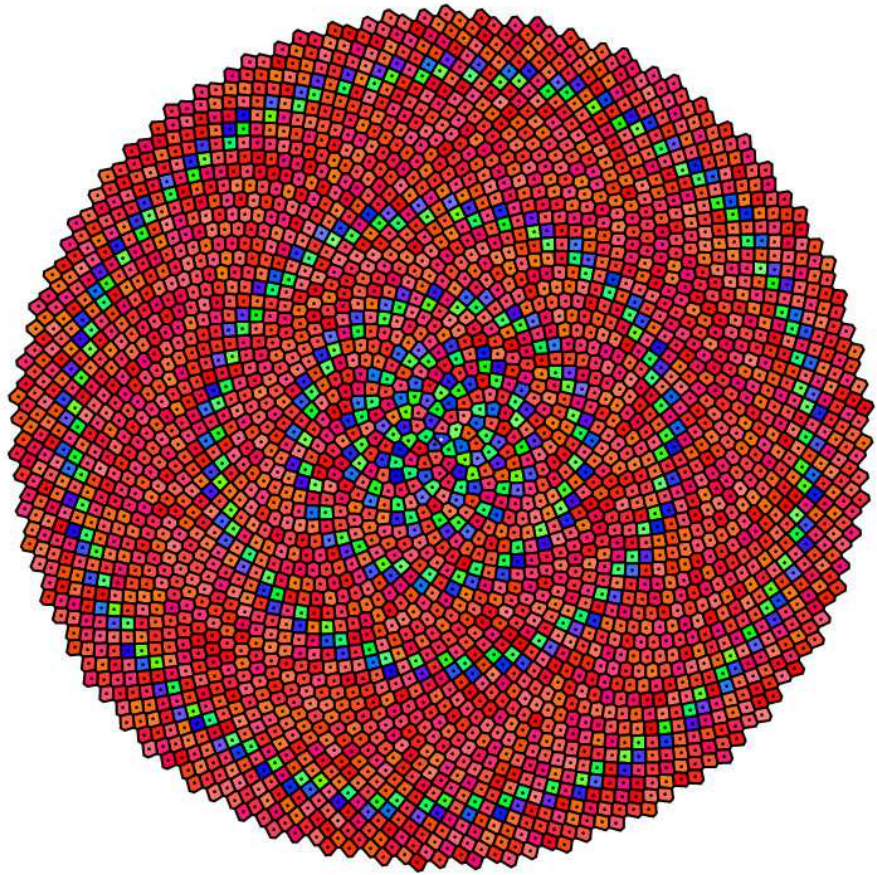}

\caption{ A set of $n=3000$ points organized on the plane according to the algorithm of phyllotaxis with the golden ratio. Each point is surrounded by its polygonal Voronoi cell whose number of sides corresponds to the number of first neighbours around the point. Blue, red and green cells are respectively pentagons, hexagons and heptagons. The three spirals joining the first neighbours of the points are called parastichies. The white dot marks the origin point $s=0$.}
\label{f1}
\end{figure}

A phyllotactic distribution of points indexed by $s$ is described by an algorithm such that the position of point $s$ is given by its polar coordinates:
\begin{eqnarray}
\rho(s)=a \sqrt{s} \textrm{~and~}   \theta(s)=2 \pi \lambda s \label{equa1}
\end{eqnarray}
which is the equation of a Fermat spiral called  generative spiral hereafter.
The parameter $a$ defines the metric scale.
Sites, indexed by positive integers $s$,  are placed on this spiral\footnote{In this paper, the first point correspond
to $s=0$ and the next, to successive integers. Taking $s\in \mathbb{Z}+1/2$ leads to a more regular structure in the core of the phyllotaxis and it appears as a $\pi/\tau$ rotation for large $s$.}, so that the azimuth between two successive points
 varies by $2 \pi \lambda $.
The best packing efficiency is obtained for  $\lambda=1/\tau$ where $\tau$ is the golden ratio $(1+\sqrt{5})/2$. The Voronoi cell of point $s$ is defined as the region of space nearer to it than to any other point of the pattern. The phyllotactic pattern of a set of 3000 points with their Voronoi cells is shown in figure~\ref{f1}.

In such a pattern, pentagons and heptagons are topological defects distributed among hexagons. These defects appear concentrated in narrow circular rings with constant width, separating large annuli of hexagons whose width increases from the core towards the periphery. In the narrow rings, pentagons and heptagons are paired as dipoles separated by hexagons whose shape is that of a square with two adjacent corners cut off. The defect rings are indeed grain boundaries separating grains of hexagonal cells, each dipole acting as a dislocation that introduces a new parastichy (the lines joining first neighbours) to maintain the density as constant as possible. The underlying arithmetic of this organization is that of the Fibonacci sequence, $f_{u}=f_{u-1}+f_{u-2}$ from $f_0=0$ and $f_1=1$, as  summarized in table~\ref{tab1}. So cells are organized in concentric blocks, grains separated by grain boundaries, surrounding a central core containing about ten points and showing an apparent disorder recalling the structure of confined two-dimensions foams. In this paper we are interested in large structures, so that
we shall consider what happens outside the core: A succession of
large grains of hexagonal cells  that are concentric circular annuli, bounded and separated by circular grain
boundaries $(f_{u-1}, f_{u-2}, f_{u-1})$ made of $f_{u-1}$ heptagonal cells, $f_{u-2}$ hexagonal cells
and $f_{u-1}$ pentagonal cells.

%
These grain boundaries serve as natural boundaries for our optimal packing problem. Outwards packing begins with the
first complete grain boundary $(13, 8, 13)$ with 13 heptagons, 8 hexagons and 13 pentagons. The core is bounded by
the 8 pentagons of the (first) incomplete grain boundary $(3, 5, 8)$. It has 3 heptagons instead of the full 8. The
additional pentagons (nearly) fulfil the topological requirement that a tiled circular domain should have a topological charge of 6
(i.e. 6 additional pentagons, a sphere having a topological charge 12)~\cite{rivier2,rivier3}.
This solves the packing efficiency problem: one grain boundary constitutes a perfect circular boundary for the domain into
which objects are to be packed.

\begin{table}
\caption{\label{tab1} a) 
Cell types in the successive rings. The number $s$ is that of each point on the generative spiral, the first neighbors of point $s$ have numbers $s+\delta s$ where $\delta s$ are all Fibonacci numbers except for $s=0$.
The large hexagonal rings or grains are labeled by the rank $u$ of the Fibonacci number $f_u$ corresponding to the medium value of $\delta s$ in a ring, the  $\delta s$  are also the numbers of parastichies of each type in the ring. The narrow rings or grain boundaries are marked by   $\|$.
b) Here the core in a) has been cleared up by starting the numbering from $s=1$ (eliminating cell $s=0$ by removing the boundary $0/2$) and making two neighbour flips to remove the $13$ parastichy from the innermost 3 cells. All $\delta s$ are Fibonacci numbers \cite{core}.}%
\begin{tabular*}{\textwidth}{@{}l*{15}{@{\extracolsep{0pt plus
12pt}}l}}
 \hline
  $u$& cell type&number of cells&$s$ from &to&neighbour separations  $\delta s$\\
 \hline
 a)\\
&pentagon&2&0&1&1,2,3,4,5 or -1,2,3,5,8\\ &hexagon&1&2&2&-2,2,3,5,8,13\\ &heptagon&3&3&5&(-3,-2,2) or (-4,-3,-2) or
(-5,-3,-2),3,5,8,13\\ &hexagon&1&6&6&-5,-3,3,5,8,13\\ &pentagon&2&7&8&-5,-3,5,8,13\\ &hexagon&1&9&9&-8,-5,-3,5,8,13\\
$\|$&hexagon&5&10&14&-8,-5,5,8,13,21\\ $\|$&heptagon&3&15&17&-13,-8,-5,5,8,13,21\\
$\|$&hexagon&5&18&22&-13,-8,-5,8,13,21\\ $\|$&pentagon&8&23&30&-13,-8,8,13,21\\ \hline i-1&hexagon& & &  &$-f_{i},-
f_{i-1},- f_{i-2}, f_{i-2}, f_{i-1}, f_{i}$\\ $\|$&heptagon&$f_{i-1}$ & & &$-f_{i},- f_{i-1},- f_{i-2}, f_{i-2}, f_{i-1},
f_{i}, f_{i+1}$\\ $\|$&hexagon&$f_{i-2}$ & & &$-f_{i},- f_{i-1},- f_{i-2}, f_{i-1}, f_{i}, f_{i+1}$\\
$\|$&pentagon&$f_{i-1}$ & & &$-f_{i},- f_{i-1}, f_{i-1}, f_{i}, f_{i+1}$\\ \hline
b)\\
&pentagon&1&1&1&1,2,3,5,8 \\ &hexagon&2&2&3&-1,1,2,3,5,8 or -2,-1,2,3,5,8\\ &hexagon&2&4&5&-3,-2,3,5,8,13\\ &hexagon&3&6&7&-5,-3,3,5,8,13\\
&pentagon&1&8&8&-5,-3,5,8,13\\   
&hexagon&1&9&9&-8,-5,-3,5,8,13\\   
$\|$&heptagon&1&10&10&-8,-5,-3,5,8,13,21\\
$\|$&hexagon&5&11&15&-8,-5,5,8,13,21\\
$\|$&heptagon&2&16&17&-13,-8,-5,5,8,13,21\\
$\cdots$ \\
 \hline
\end{tabular*}
\end{table}

\subsection{Defect rings}

In the middle
 of hexagonal grains it is easy to identify three parastichies which are spirals running through neighbouring  points. The %
three curves intersect with angle close to $2\pi/3$, but when the spirals approach 
the grain boundary domain only two remain apparent.  The third disappears to restart in an orthogonal direction through the truncated square hexagons of the grain boundary. The two apparent parastichies are orthogonal and give the directions of the dipoles. The Voronoi cells on the grain boundaries are nearly squares. The two apparent parastichies are orthogonal to the square edges, the third runs  along one diagonal only and there is a flip of the diagonal across the hexagons of the grain boundary.

The elementary dipoles (in which heptagon and pentagon are nearest neighbours) %
are oriented along parastichies $f_u$
refereed by $u$ in table~\ref{tab1} but   there is a flip %
of orientation  from one grain boundary to the next. %
Nevertheless all dipoles %
make the same angle  $\textrm{arccot}(f_u/f_{u-1})\simeq \textrm{arccot}(\tau)$ in absolute value
with the radial direction. This results
simply from the description of grain boundaries given in appendix~A as
a strip in a square lattice.

The perimeter of a ring of dipoles is determined by its number of dipoles and hexagonal cells constituting it. Table~\ref{tab1} shows that this number belongs to the Fibonacci sequence %
so that the ratio of the radii of two successive rings of dipoles is approximated by the golden ratio $\tau$. Strictly, as shown %
in appendix A, this ratio is
$\sqrt{f_{2u+3}/f_{2u+1}}$ converging to    $\tau$ for large Fibonacci numbers.
It can be seen on figure~\ref{f1} that the dipoles occur as     singletons, or in pairs, %
distributed on the
ring according to an approximant of a quasicrystalline sequence \cite{rivier6},  as explained in appendix~A.%
When moving from one ring to the next, away from the core,
 the evolution of the sequence is determined by the Fibonacci %
 inflation/deflation rule where a singleton becomes a pair and a pair becomes a pair plus a singleton.

\subsection{Metric properties}

\begin{figure}[tpb]
\caption{(a) Variation of the area of Voronoi cells for a pattern of $n=3000$ cells. This is obtained with the scaling parameter $a=1$ in equation~\ref{equa1}  leading to an area close to $\pi$. This area tends toward $\pi$, but with rapid variations  in grain boundaries. Inserted  is an histogram of the area of cells, notice that the scale is very inflated.(b) First neighbour distances between points $s$ and $s+\delta s$. Blue, for the interval $\delta s$ equal to the
smaller positive Fibonacci number in the list of table 1. Green, for the next interval, red for the third positive
interval and purple for the last one (occurring only if the  Voronoi cell is an heptagon). So  green, blue and red
correspond to distance along the three visible parastichies. Each continuous curve corresponds to a given Fibonacci
number that appears in different rings. For instance, $f_{11}=55$ appears between $s=101$ and $2254$ leading to a
continuous curve which is successively purple, red, green and blue. Lower and upper crossings of two curves corresponds
to a grain boundary, other crossings are in the middle of a hexagonal grain. An histogram of all distances is inserted.
The hook for distances close to $1.9$ and $2.3$ is an artefact due to the finite number of points.
The distances lies in the range $[(2\pi/\sqrt{5})^{1/2}, (2\pi)^{1/2}]$. }
\includegraphics{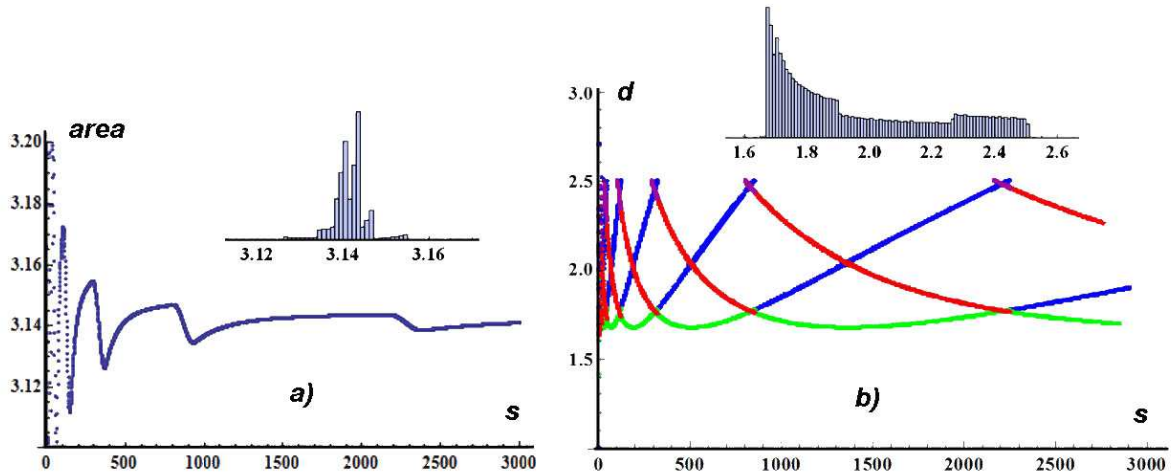}

\label{f2}
\end{figure}

The area of a disk %
   of radius $\rho$ which contains $s$ points is $\pi \rho^2=\pi a^2 s$ so that the average %
    area per point has the value  $\pi a^2$. The %
Voronoi decomposition is a tessellation which attributes %
 a specific area to each point. But  the %
 areas of %
 Voronoi cells are not all identical %
 as shown in figure~\ref{f2}a even if fluctuations are small. The area of  a Voronoi cell  falls
 rapidly each time a ring of dipoles is crossed as $s$ increases and it %
 tends towards $\pi a^2$ for large $s$. The fluctuations are small %
 immediately outside the core of the pattern. The histogram %
  describes the level of homogeneity attained with phyllotaxis. The standard deviations of similar histograms are $\delta=0.01589$ for
   $n=6000$, $\delta=0.02246$ for
    $n=3000$ and $\delta=0.03171$ for
     $n=1500$ (assuming $a=1$), rapidly decreasing with size $n$.

 %
The behaviour of the distances  between first neighbour points is shown on  figure~\ref{f2}b. In grain boundaries where Voronoi cells are
slightly deformed squares, the cell area is approximatively $\pi a^2$ so that the distance between points, along a square
edge, is  $\sqrt{\pi}\simeq1.772$ (with $a=1$) and along a diagonal, is $\sqrt{2\pi}\simeq2.506$. In the grains
where Voronoi cells are more clearly hexagons  two distances are close to $\sqrt{\frac{3\pi}{\sqrt{5}}}\simeq2.053$.   The  distance between the two points defined by $s$ and $s+f_u$  as a function of $s$ is obtained using the
relation (see \cite{yeatts,sadocriviercharvolin}):
 \begin{equation}
d_u(s)=f_u (\frac{1}{4s}+\frac{ s (-2\pi f_{u-1}+2\pi \tau^{-1}f_u)^2}{f_u^2})^{1/2}, \label{equa2}
\end{equation}
this relation\footnote{This relation suppose  that $\delta s$ is infinitesimal compared to $s$. An estimation of the
accuracy can be obtained comparing the distances computed between points $s$ and $s+\delta s$ with that between $s+\delta
s$ and $s$. They must be equal but computed values are slightly different. An average between both is very accurate.}  fits the numerical values given on figure~\ref{f2}b. It can be checked that minimal values are
\begin{equation}
\sqrt{2 \pi } f_u \sqrt{|\frac{1}{\tau }-\frac{f_{u-1}}{f_u}|} \label{equa3}
\end{equation}
all close to and converging toward $\sqrt{\frac{2\pi}{\sqrt{5} }}\simeq1.67$.
All distances between first neighbours are in the range $[\sqrt{\frac{2\pi}{\sqrt{5} }},\sqrt{2\pi}]$, whatever will be $s$ and $u$.
This property, specific to the choice of $\lambda=1/\tau$
is very important to ensure the best uniformity in all orientations.
The reason is that $\tau$  is approximated by successive truncations of its continuous fraction  (having only $1$) which are the ratio of two successive Fibonacci numbers and so converging smoothly~\cite{rivier4,adler98}. It is this property which leads to have minimal values for distances given by the equation~\ref{equa3} converging toward a finite value.
Other ratios (with a tail of their continuous fraction expansion containing other integers) introduce some distances between  neighbours decreasing rapidly with the $\delta s$ separation between neighbours.

\section{Phyllotaxis on the hyperbolic plane}
\subsection{The Poincar\'e disc}
The Poincar\'e disc model  is a simple way to represent the hyperbolic plane as presented in appendix~B.
A point in the hyperbolic plane is defined by polar coordinate $(\rho,\vartheta)$ and is represented on the Poincar\'e  disc by $(r,\vartheta)$ where lengthes are obtained with a given metric $d\sigma^2=4(dr^2+r^2d\vartheta^2)/(1-r^2)^2$. So $r=1$ represents points at infinity.
 This metric has the form of an Euclidean metric in polar coordinates divided by a function of $r$ only: it is locally an Euclidean metric.

\subsection{Pattern}
\begin{figure}[tpb]
\caption{Hyperbolic phyllotaxis represented inside the Poincar\'e disc. The pattern is obtained using equation~\ref{equa4} with 3000
points and $a=1/20$. Blue, red and green cells are respectively pentagons, hexagons and heptagons. All area and distances are scaled so that the mean area per points is $\pi$. With this choice the curvature $\kappa=-1/R^2$ is given by $R=1/a$, with the parameter $a=1/20$. By effect of the Poincar\'e representation size of cells seems to decrease going toward the limit
circle.}
\includegraphics{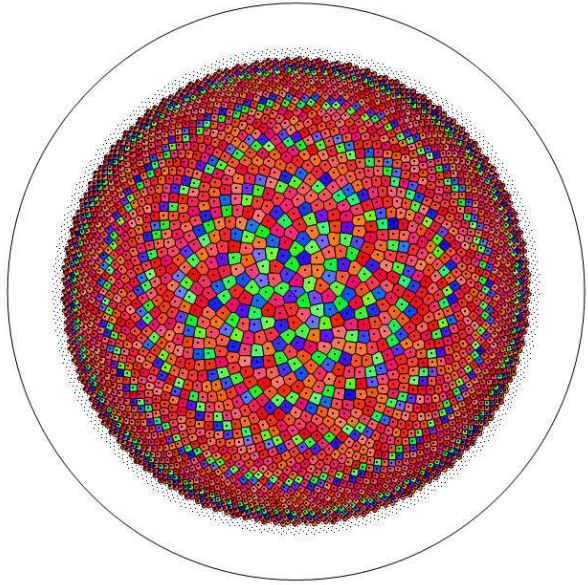}

\label{f3}
\end{figure}

As for  phyllotaxis in the Euclidean plane, a point is defined by an integer $s$ on a spiral and the pattern obtained is shown in figure~\ref{f3}.
The number of points in an hyperbolic cap of radius $\rho(s)$ is proportional to $s$. We choose  the $\rho(s)$ function so that the number of points enclosed in the domain is $\pi a^2 s$, as it is for the Euclidean plane phyllotaxis. Using the equation~\ref{equaB1} given for the area enclosed by a circle in appendix~B,  we write $2\pi a^{2}s/2=2\pi(\cosh\varphi-1)$,
with $R=1$ to have the Gaussian curvature equal to $\kappa=-1$, then $\varphi=\rho(s)$. The true radius in the hyperbolic plane is $\rho(s)=\cosh ^{-1}\left(\frac{ a^2 s}{2} +1\right)$. This radius is a length measured on the hyperbolic plane,
but on the  Poincar\'e disc, The Euclidean distance from the origin to the representation of a point   $r=\tanh(\varphi/2)$ is given by
\begin{eqnarray}
r(s)=\tanh \left(\frac{1}{2} \cosh ^{-1}\left(\frac{ a^2 s}{2}+1\right)\right). \label{equa4}
\end{eqnarray}
The factor $a^2/2$ relating $s$ and $\rho(s)$ allows to have the radius in the hyperbolic plane $\rho(s)\simeq a \sqrt{s}$ for small $s$ as in flat phyllotaxis.
The spiral equation in the hyperbolic plane is $(\rho(s)\cos(2\pi \lambda s),\rho(s) \sin(2\pi \lambda s))$ where $\lambda$ is a parameter exactly similar to the $\lambda$ parameter introduced in plane phyllotaxis. Similarly the true hyperbolic phyllotaxis corresponds to the choice $\lambda=1/\tau$.
 As shown in 3.4, the  $\rho(s)$ behaviour and the choice of $\lambda=1/\tau$ ensure the best homogeneity and isotropy for the
distribution of points.

%
%
\subsection{Rings of dipoles}
A Voronoi decomposition of the phyllotaxis mapped on the Poincar\'e disc can be obtained as if the structure in this disc
was Euclidean. This is  justified by the fact that the metric is locally Euclidean. If two
points are neighbours with the hyperbolic metric they are also neighbours  using the local Euclidean metric on the Poincar\'e map
\footnote{More precisely, this is also related to the fact that two triangular Delaunay decompositions obtained using the Euclidean metric of the plane representation or the hyperbolic metric are formed with the same set of triangles. These sets of triangles are defined by the  circumcircles of triangles selected so that they do not enclosed any points of the phyllotaxis.  Because circles in the hyperbolic plane are mapped onto circles in the Poincar\'e representation the two selected sets are identical. Consequently neighbouring relations are the same with the two metrics.}.
On  figure~\ref{f3}, Voronoi cell edges are straight segments; on an exact Voronoi decomposition they must be arc of circles representing geodesic lines of the Poincar\'e disc but as these segments are very short, the approximation is clearly good.

As in plane  phyllotaxis it is possible to define hexagonal grains separated by grain boundaries formed with pentagon-heptagon dipoles. It is the change in the list of separations $\delta s={f_{u-1},f_u,f_{u+1}}$ between neighbours in a grain and in the following one which govern a grain boundary. So the description given in appendix A remains the same, still related to Fibonacci numbers. Grain boundaries are indeed the same in the three examples of phyllotaxis.

%
%
\begin{figure}[tpb]
\caption{Area and distances on the hyperbolic plane. All area and distances are scaled so that the mean area per points is $\pi$. With this choice the curvature $\kappa=-1/R^2$ is given by $R=1/a$, with the parameter $a=1/40$. (a) Area of Voronoi cells for the  hyperbolic phyllotaxis.   The area are close to the
mean value $\pi$, strong variations appear near rings of defects. Nevertheless these strong variations decrease
as the area tends toward $\pi$ for large $s$. (b) Distance  between point $s$ and its first neighbours $s+\delta s$ with positive $\delta s$.  Green, blue and red
correspond to distance along the three visible parastichies. Distances are confined between the limits $\sqrt{2 \pi}$ and
$\sqrt{2\pi/\sqrt{5}}$.
}
\includegraphics{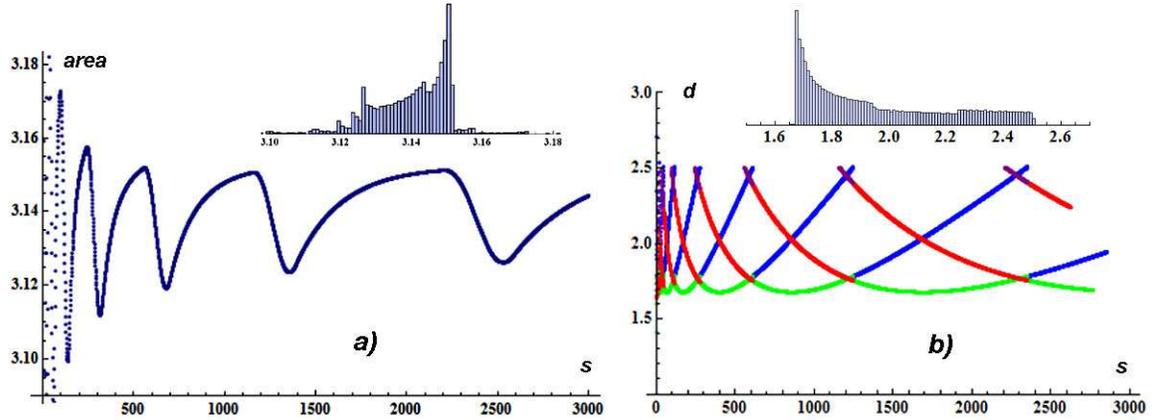}
\label{f4}
\end{figure}

%

\subsection{Metric properties in hyperbolic phyllotaxis}

The $\rho(s)$ function has been chosen so that the area enclosed in a circle of radius $\rho$ is $\pi a^2 s$ on a hyperbolic plane of $\kappa=-1$ Gaussian curvature. On  figure~\ref{f4}a the evolution of the  area per cell is shown scaled in order to have the mean area equal to $\pi$ as in the plane case; so the curvature is $\kappa=-1/R^2$ with $R=1/a$. Area fluctuations are stronger, with a smaller damping than in the plane case.

It is possible to get distances between first neighbour points from the coordinates; this is presented on figure~\ref{f4}b but it is also possible to have an accurate analytical relation allowing to follow behaviour of distances with $s$.
On a plane representation,  the distance, measured on the plane, separating two close points whose indices are $s$ and $s+\delta s$ with $\delta
s=f_u$ is given with a good accuracy by:
\begin{equation}
(d_u(s))^2=f_u^2 [r^{\prime}(s)+(\gamma_u/f_u)^2 r(s)^2], \label{equa5}
\end{equation}
with $\gamma_u=(-2\pi f_{u-1}+2\pi \tau^{-1}f_u)$  and $r(s)$ given by equation~\ref{equa4}.
Equation~\ref{equa2} given for a plane phyllotaxis is derived from this equation (see the foot note in 2.3 on the accuracy of
this relation),
 but  to have the true hyperbolic  distances  requires to take account of a correcting factor to the
metric. This  factor is $ 1 /(1-r(s)^2)$. Then the distances are given by:
\begin{equation}
d_u(s)=f_u \frac{1}{1- r(s)^2}(\frac{-a^2}{4s( 1+s a^2)^3}+\frac{ r(s)^2 (-2\pi f_{u-1}+2\pi
\tau^{-1}f_u)^2}{f_u^2})^{1/2}. \label{equa6}
\end{equation}
This relation fits the curved given on  figure~\ref{f4}b after normalization by a scaling factor $1/a$ in order to have mean area of cells equal $\pi$.

As the sequence of rings of dipoles are the same whatever the curvature as developed in appendix~A, the
effect of the negative curvature is to decrease the width of hexagonal grains which tends to be a constant for large $s$. Effectively, the radius of a grain boundary is  $\sinh \varphi_u=\sqrt{f_{2u+1}/\pi}/2$ which leads for large $\varphi_u$ to $\varphi_u\simeq \ln(\sqrt{f_{2u+1}/\pi})$. Two successive grain boundaries defined by $u-1$ and $u$ give a width $\varphi_u-\varphi_{u-1}\simeq\ln(\sqrt{f_{2u+1}/f_{2u-1}})$ very close to the constant $\ln \tau$.

Like in the case of the plane, the maximum for first neighbour distances is still $\sqrt{2 \pi}\simeq 2.506$ corresponding,  to cells close to squares with a distance corresponding to the diagonal. The minimal values of $d_u(s)$ (equation~\ref{equa6}) are $\sqrt{2 \pi } f_u \sqrt{|\frac{1}{\tau }-\frac{f_{u-1}}{f_u}|}$ very close to and converging toward $\sqrt{\frac{2\pi}{\sqrt{5} }}\simeq1.67$. Distances are confined in the same domain as for the plane phyllotaxis, a consequence of the choice of the parameter $\lambda=1/\tau$.

%
%
  %
\begin{figure}[tpb]
 \caption{Phyllotaxis on a sphere with 4001 points.  Blue, red and green cells are respectively pentagons, hexagons and heptagons; the white point is a pole. Ring of dipoles appear clearly. Near the equator there is a large hexagonal domain even if cells look like squares, in fact they have six edges two of which being very small.The structure has a chiral symmetry around the axis joining the two poles. }
\includegraphics{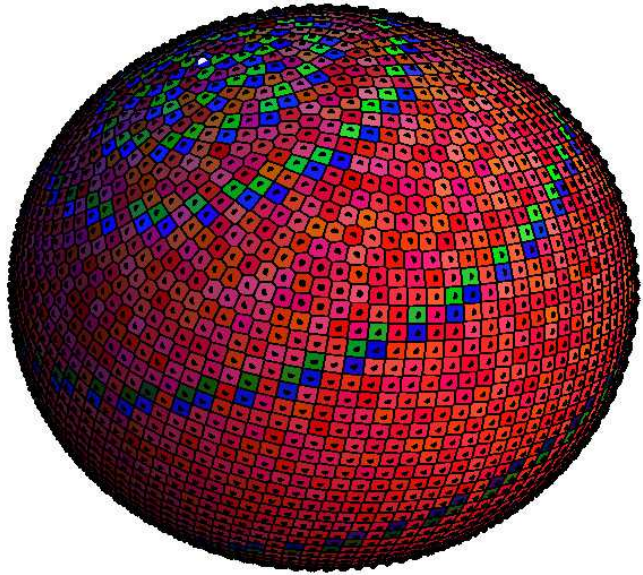}

\label{f5}
\end{figure}
%
%
\section{Phyllotaxis on the sphere}
\subsection{Pattern}
The algorithm used to build the phyllotactic configurations shown in figure~\ref{f5} is such that the position of point $s$
is given by the spherical coordinates $(\rho,\phi,\theta)$.
These coordinates are
$\rho=R$ the sphere radius,  $\theta= 2 \pi \lambda s$, the azimuthal angle and $\phi=\arcsin( s^\prime/ \nu)-\pi/2$, the polar angle.  The total number   of points on the sphere  is $n=2\nu+1$. With this choice, the integer $s^\prime$ goes from $-\nu$ to $\nu$ with $s^\prime=-\nu$ or $s^\prime=\nu$ for points on poles, and $s^\prime=0$ for a point on the equator.  For comparison with plane phyllotaxis it could be helpful to have $s=0$ on the north pole using $s=s^\prime+\nu$.

Spherical phyllotaxis is obtained by mapping of points on a finite cylinder which is tangent to the sphere along the equator, so which has the same radius $R$ and whose finite hight is $2R$, the area of this finite cylinder is  $4\pi R^2$ as that of the sphere.
The points on the cylinder  are located on a perfect helix defined by $\theta= 2 \pi \lambda s$ and a pitch related to the number of points on the sphere. They can be consider as  a perfect crystal without
defect wrapped on the cylinder, so with a constant  density of points. This is an example of phyllotaxis on a cylinder as described in \cite{coxeter1,rothen}. With $\lambda=1/\tau$, a point $s$ has six neighbours at position $s\pm\delta s$ with $\delta s$ equal to three successive Fibonacci numbers, depending on the pitch of the helix drawn on the cylinder. These points are mapped on the sphere orthogonally to the polar axis and so the area per points on the sphere is the same as on the cylinder. This results from the property of the
area of a spherical zone enclosed on a sphere of radius $R$, between two parallel planes at distance $h$ to be the
same as the area of a finite part of a cylinder of radius $R$ enclosed between the two same parallel planes
orthogonal to the cylinder axis. This justify the choice of the polar angle on the sphere $\phi=\arcsin( s^\prime/ \nu)-\pi/2$ deduced from the cylindrical coordinate $(R, \theta, R \sin(\phi +\pi/2))$.
But the projection on the sphere, even if it conserves area, introduces inhomogeneous shearing in the structure. This shearing, at constant area, is given by a compression factor $\cos(\phi)$ changing the equator length into a parallel length (at $\phi$), and an expansion factor $1/\cos(\phi)$ along meridians.
This shearing could change the incidence relations between neighbours so first neighbours are not necessarily the same as on the ``crystalline'' cylinder and defects appear.
The area associated to cells is close to be constant as in plane phyllotaxis. The choice of $R$ is
arbitrary but in order to have cell area close to $\pi$ as in plane phyllotaxis  $R=\sqrt{2\nu+1}/2$.

A Voronoi decomposition of the set of points on the sphere reveals the phyllotaxis and how defects appear \footnote{
The Voronoi decomposition of a set of points on a sphere has been done using a 3D Voronoi analysis considering the set
of points on the sphere to which is added the center of the sphere. Then Delaunay tetrahedra are obtained. They are formed with Delaunay triangles on the sphere surface completed by the center of the sphere. The Voronoi decomposition on the surface is deduced from this Delaunay triangulation of the surface.}.
As in plane phyllotaxis the Voronoi cells  are hexagons, pentagons or heptagons,
with most of the Voronoi cells being hexagons.  We call  the large domains with only hexagonal Voronoi cells in hexagonal grains,  even if they are not regular, and  the circular distribution of heptagonal, hexagonal and pentagonal cells are grain boundaries.

\subsection{Rings of dipoles}

\begin{figure}[tpb]
\caption{Evolution of rings of defects near the equator increasing the number $n=2 \nu+1$ of points on the sphere. Colours
of Voronoi cells are red for hexagons, green for heptagons and blue for pentagons. Appearing on the equator they are two
interpenetrating rings. Then some cells having only four sides appear in yellow.}
\includegraphics{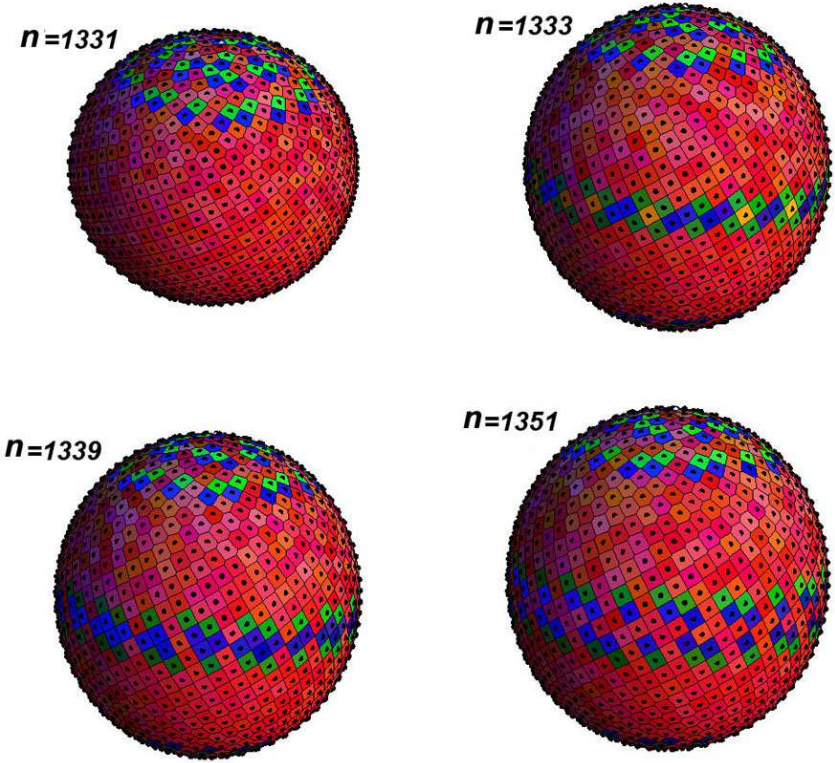}

\label{f6}
\end{figure}

The number of rings of dipoles depends on the number $n$ of points on the sphere, related to its radius $R=\sqrt{n}/2$.  Figure~\ref{f6} shows how new dipoles appear increasing $n$.
The evolution of rings of dipoles  is related to the most important fact that the distributions of dipoles and
hexagonal cells along these rings follow the  one dimensional sequences approximant of quasicrystals as developed in appendix A.  The  number of cells in these rings  is   $2f_u+f_{u-1}$, where $f_u$ is the number of dipoles, and
$f_{u-1}$, the number of hexagons. The length of circles defining rings of defects are the length of the strip in a square lattice, so this length is the
same in plane and spherical phyllotaxis. It is $\sqrt{(f_u^2+f_{u+1}^2)\pi} $ or $ \sqrt{f_{2u+1}\pi}$ if we suppose an
area of cells given by $\pi$ and then a cell edge of $\sqrt{\pi}$ when cells are squares. Here and in the following we set the sphere radius $R=\sqrt{2\nu+1}/2$ in order to  have average cell area equal to $\pi$ as in plane phyllotaxis. This is then used to estimate the number of cells in hexagonal grains.

On a sphere of radius $R=\sqrt{2\nu+1}/2$ containing $n=2\nu+1$ sites, a circle defined by the polar angle $\phi$  border
a spherical cap containing   $\nu(1-\cos \phi)$ sites.  If the circle corresponds to a grain boundary of length
$\sqrt{f_{2u+1}\pi}$ this leads to a polar angle $\phi$ given by $\sin \phi=\sqrt{\frac{f_{2u+1}}{(2\nu+1)\pi}}$. We
estimate the enclose number of points inside the spherical cap bordered by the grain boundary to be:
$\nu(1-\sqrt{1-\frac{f_{2u+1}}{(2\nu+1)\pi}})$. This is an estimation, in order to get size of hexagonal grains we have to take
account of the width of the grain boundary and suppose that the estimation corresponds to the medium of the $f_{u-1}$
hexagonal cells in the grain boundary. An estimation for grain boundaries bounds, sometime shifted from $1$ is:
\begin{equation}
 {\lfloor (5-f_{u-1})/2+\nu(1-\sqrt{1-\frac{f_{2u+1}}{(2\nu+1)\pi}})\rfloor,\lfloor
 (f_{u-1}+3)/2+\nu(1-\sqrt{1-\frac{f_{2u+1}}{(2\nu+1)\pi}})\rfloor}.
\label{equa7}
\end{equation}

Increasing the number $n$ of points on the sphere, and so decreasing the curvature, new rings of dipoles (grain boundaries) appear on the equator $\phi=\pi/2$. These is presented on  figure~\ref{f6} in the range $n\in[1331,1351]$.
New grain boundaries appear by pair: one in each hemisphere. As they have a finite width just after the threshold of appearance the two grain boundaries are interpenetrating; this leads to some cells which could have four edges.
Setting   $\sin\phi=1$  gives the the threshold for the number of points on the sphere for
which a new grain boundary appears on the equator. The first set of these values are $n=(1, 2, 4, 11, 28, 74, 194,$
$508, 1331, 3484, 9122, 23881,\ldots)$.
When a new grain boundary, related to the Fibonacci number $f_{2u+1}$, appears the previous one, related to the Fibonacci number $f_{2u-1}$, enters a polar cap of angle  $\phi$ given by $\sin \phi=\sqrt{\frac{f_{2u-1}}{n \pi}}$ or
$\sin \phi=\sqrt{\frac{f_{2u-1}}{f_{2u+1}}}$. It results that $\phi\simeq \arcsin \tau^{-1}$ about $0.666$ $\textrm{rd}$.
This value of $\phi$  does not depend on $n$ as soon the approximation $\sqrt{\frac{f_{2u-1}}{f_{2u+1}}}\simeq\tau^{-1}$ is good.

%
%
%

\begin{figure}[tpb]
\caption{a) Evolution of the polar angle $\phi$ of grain boundaries with the number $n$ of points on spheres (logarithmic horizontal scale). Each curve corresponds to grain boundaries  defined by $u=5,6,7,8,9,10,11,12$. For a given number $n$ (or a given radius $R=\sqrt{n}/2$) there is a grain boundary near the equator if there is a curve such that $\phi\simeq \pi/2$. The two black points refer to the two examples of spheres whose cells area and distances are given on figure~\ref{f8}. b) Evolution, with the Gaussian curvature, of the arc length  which is the radius of grain boundaries (for instance $R \phi$ in spherical case). There is a limit to these curves in spherical case corresponding to the appearance of  grain boundaries at the equator. }
\includegraphics{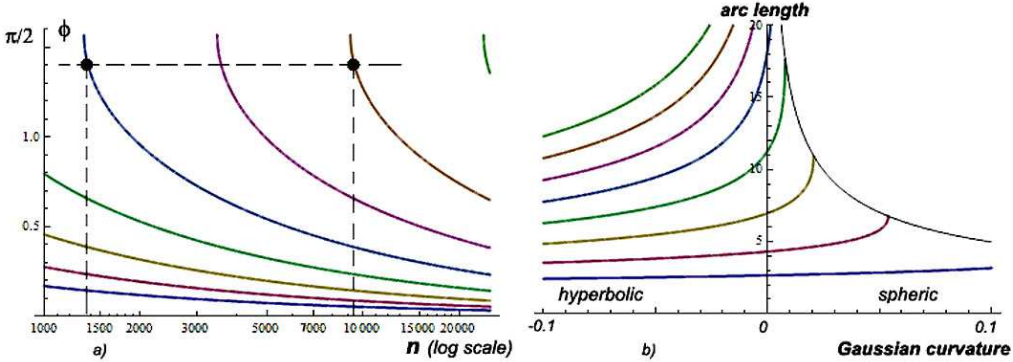}

\label{f7}
\end{figure}

The same evolution repeats itself quite regularly at each appearance of a new grain boundary as shown on figure~\ref{f7}. Using a semi-logarithmic horizontal scale, the curves can be translated from one to the next.

%
%
%
%

\subsection{Metric properties on the sphere}

\begin{figure}[tpb]
\caption{ (Left) Area of Voronoi cells for the spherical phyllotaxis.
All area and distances are scaled so that the mean area per points is $\pi$. Two examples are given with $n=9301$ and $n=1351$  points on the sphere.
(Right) For the same spheres, distance  between points $s$ and its first neighbours $s+\delta s$ with positive $\delta s$ in the north hemisphere and negative $\delta s$ in the south.  Green, blue and red
correspond to distance along these three visible parastichies. Distances are in the limits $\sqrt{2 \pi}$ and
$\sqrt{2\pi/\sqrt{5}}$. Curves for the two examples $n=9301$ and $1351$ look very similar, mainly for distances: this is because the $n$ values  are chosen to have grain boundaries at the same angular positions as given on figure~\ref{f7}.}

\includegraphics{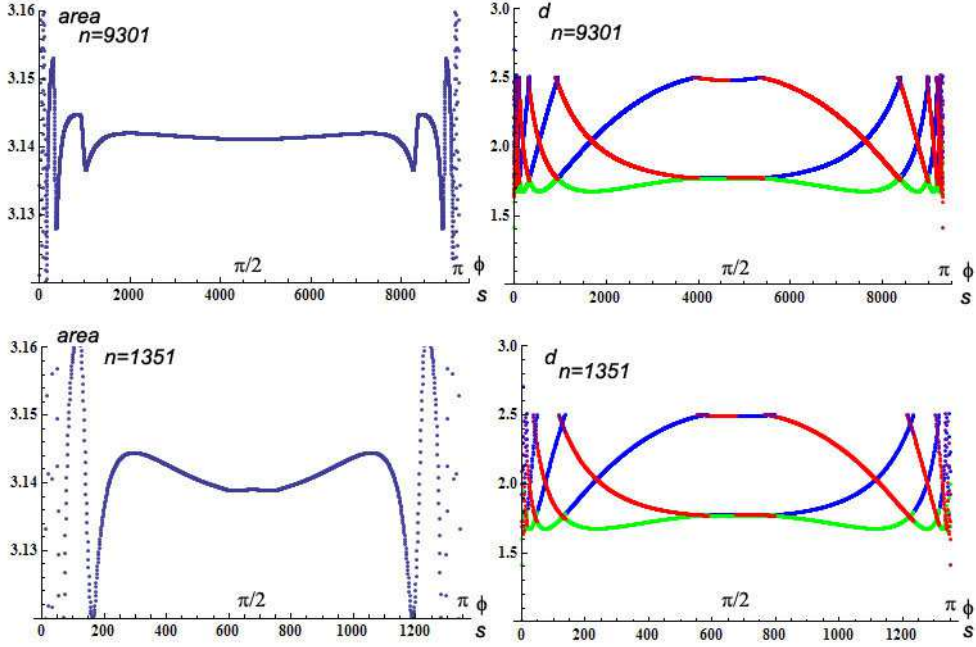}

\label{f8}
\end{figure}

Figure~\ref{f8}  shows that, each time the number $n$ of point on the sphere is translated from one value to the next along the logarithmic scale of figure~\ref{f7}, the distances between first neighbour points are reproduced while a new grain boundary is introduced. The local order varies within the same limits whatever the number of points.
Far away from grain boundaries the area are close to the mean value $\pi$ corresponding the the choice $R=\sqrt{n}/2$. There are  strong variations crossing   grain boundaries near the poles which decrease rapidly.

In hyperbolic case we have evaluated distances in  the Poincar\'e representation using a given metric. It is convenient to use a similar method for the spherical space using a representation of the sphere as described in appendix~C, which is no more than a stereographic projection of the sphere on a tangent plane.
On this plane representation of a spherical phyllotaxis  the distance separating two close points whose indices are $s$ and $s+\delta s$ with $\delta s=f_u$ is always given by the equation~\ref{equa5}, but where $r(s)$ is given by the equation~\ref{equaC1}. With equation~\ref{equa5} this distance is evaluated with the plane metric, so to have the true distances on the sphere
needs to take account of a correcting factor to the metric  which is $ 1 /(1+r(s)^2)$ as given in appendix~C by equation~\ref{equaC2}. Then the distances are given by:
\begin{equation}
d_u(s)=f_u \frac{1}{1+ r(s)^2}(\frac{-a^2}{4s(- 1+s a^2)^3}+\frac{ r(s)^2 (-2\pi f_{u-1}+2\pi
\tau^{-1}f_u)^2}{f_u^2})^{1/2} \label{equa9}
\end{equation}
This analytical distances fit the distances presented on the  figure~\ref{f8} obtained from coordinates of the spherical phyllotaxis taking account of a scaling factor $1/a$ in order to have mean area of cells equal $\pi$. The important conclusion is that these distances are again confined in the same domain found for the plane phyllotaxis. Like in the case of the plane, the maximum for distance is still $\sqrt{2 \pi}\simeq 2.506$ corresponding,  to cells close to squares with a distance corresponding to the diagonal. It can be checked that minimal values of $d_u(s)$ (equation~\ref{equa9}) are $\sqrt{2 \pi } f_u \sqrt{|\frac{1}{\tau }-\frac{f_{u-1}}{f_u}|}$ very close to and converging toward $\sqrt{\frac{2\pi}{\sqrt{5} }}\simeq1.67$ like for the plane phyllotaxis.
As in the plane and hyperbolic cases, it is this confinement of distances which is the signature of the best uniformity in all orientations.

\section{Conclusion}

The spiral organization generated by the algorithm of phyllotaxis with the golden ratio ensures the best packing efficiency of points on a plane in a situation of circular symmetry \cite{ridley}. We examined here how this simple schema could be affected when the bearing surface presents a positive or negative Gaussian curvature, limiting ourselves to curvature radii much larger than the mean distance between points.

Drawings of figures~\ref{f1},~\ref{f3} and~\ref{f5} show that the phyllotactic patterns always present the same appearance. They are all made of a core, without any obvious discernable order,
surrounded by an ordered series of alternate concentric rings. The cores are the same, they occupy a limited region around the centres of the patterns where the curved surfaces can be assimilated to their tangent plane. The alternate concentric rings are respectively large rings, or grains, containing points with six first neighbors and narrow rings, or grain boundaries, containing also points with five and seven first neighbors associated in dipoles, or dislocations. Those topological defects are needed to maintain the density as constant as possible in this situation of circular symmetry.

A detailed analysis of grain boundaries makes apparent the fact that they can be found identical to themselves on the three kinds of surfaces. The number, distribution and orientation of dipoles on each of them, as well as their evolution from one to the next, do not depend on the curvature of the surface, their perimeters just follow the Fibonacci series as shown in table I. This directly proceeds from the mathematical structure of the algorithm of phyllotaxis with the golden ratio as developed in appendix A. Figure~\ref{fA1} shows indeed that the dipoles are organized along a grain boundary according to a quasicrystalline sequence whose evolution from one grain boundary to the next is driven by a well defined inflation/deflation rule. The distributions of the whole set of points on surfaces of different curvatures must respect this most remarkable structural invariance of the grain boundaries as developed in the following paragraphs.

As the perimeters of the grain boundaries are constrained to follow the Fibonacci series, the distance between them on the surfaces, or the widths of the grains they enclose, vary differently according to the sign of the curvature. If we call $P$ the perimeter and $R$ the radius of curvature, the radii of the grain boundaries indeed varies  as $P/2\pi$ on the plane, $R \arcsin(P/2\pi R)$ on the sphere and $R \textrm{argsh}(P/2\pi R)$ on the hyperbolic plane  and the width of a given grain decreases as the curvature goes from a positive to a negative value. This results in the more or less rapid dampening of the area per cell around $\pi$ visible on figures~\ref{f8},~\ref{f2} and~\ref{f4}.

As the dipoles on the grain boundaries have well-defined alternate orientations relative to the radius vector with an angle close to $\pi/2-\arctan(\tau)$, the anchoring conditions of the parastichies at the two limits of any grain are the same whatever the curvature of the surface and the width of the grain. This constraint manifests itself by the fact that the distances between first neighbors oscillate, along more or less extended horizontal scales, between the same limits and cross each other at the same level on the whole plane and hyperbolic plane as shown by figures~\ref{f2} and~\ref{f4}. However, on the sphere, such an oscillating behaviour is observed on a limited polar cap only and not around the equator, as shown on figure~\ref{f8}. This holds to the fact that, when two new grain boundaries merge along the equator, their dipoles are parallel and stay parallel as they move towards their pole as the sphere grows. The anchoring conditions of the parastichies in the grain surrounding the equator do not alternate as they do in other grains and the shape of its Voronoi cells evolves differently than in normal grains. Figures~\ref{f5} and~\ref{f6} show that cells in the equatorial belt may keep a shape close to that of a square rather than becoming hexagonal as in normal grains.

Finally, owing to the inflation/deflation rule determining the quasi crystalline sequences of the grain boundaries they form by themselves a self similar set which is scale invariant as the characteristic distance is changed by a factor which is an approximant of $\tau^n$. When on the plane, this self similarity is transferred onto the pattern as, assimilating grain boundaries with ideal circles, their perimeter  vary as successive approximants of $\tau$ and the area of the grains vary as $\tau^{2n}$. This is however not valid on the sphere or the hyperbolic plane where the area of the grains do not follow a variation in $\tau^{2n}$.

\section{Appendix: Grain boundaries and inflation-deflation symmetry}
  %
\begin{figure}[tpb]
 \caption{Strip cut in a square lattice. This strip is a good approximation of the ring of defects containing $55$
points. Color of the points correspond to the type of their Voronoi cells: hexagon (red), heptagon (green) and pentagon (blue).  The first point in green is down, then points
are numbered, increasing by $21$ going up or decreasing by $34$ going right. This strip can be divided into three strips
of heptagons, hexagons and pentagons. }
\includegraphics{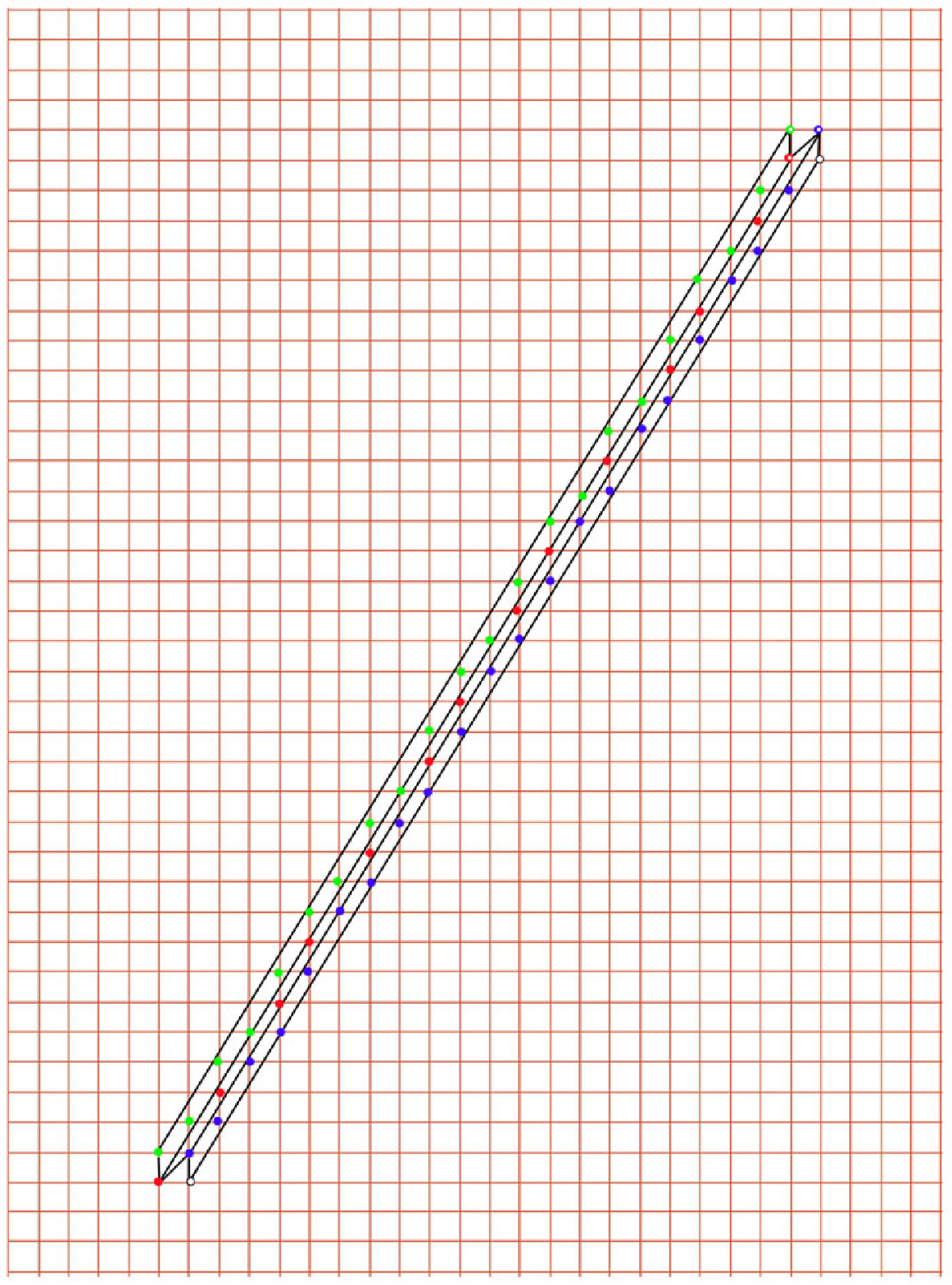}

\label{fA1}
\end{figure}
%
%
The way dislocation dipoles are organized along circles is strongly related to 1D quasicrystals  \cite{rivier6}. A Fibonacci 1D quasicrystal can be obtained using an inflation-deflation rule iteratively applied to a sequence of long and short segments. This  inflation-deflation rule is $L \rightarrow L + S$ and $S\rightarrow L$. Starting simply from a short segment this specific rule gives a quasicrystal after an infinite number of iterations, or with a given number of iterations, a finite structure with a number of short and long segments given by two successive  Fibonacci numbers.  Consider now dipoles formed by an heptagon and a pentagon in contact along grain boundaries. There are isolated dipoles, singletons and pair of close dipoles. The application of the rule changes a singleton into a pair of dipoles and changes pair of dipoles into a pair  and a singleton, and transforms one ring of dipoles into the next one. So there is an inflation-deflation symmetry associated with radial change relating defects in this structure. This can be checked counting the number of pair of dipoles or of isolated dipoles which are successive Fibonacci numbers, on circles of defects.

One dimension quasicrystals can be obtained by the cut and projection method,  selecting the points of a square lattice falling inside  a strip defined by translation of a cell of the lattice along a straight line of slope $\tau$. Approximants are obtained when the slope is a convergent of $\tau$ given by the ratio of two successive Fibonacci numbers. The figure \ref{fA1} which describes how a grain boundary can be derived from a selection of points in a square lattices is similar to the cut and projection method applied to approximants.

The description of grain boundaries as a strip of squares, allows to estimate their lengths. We suppose, as it is done in the text, a square edge length $d=\sqrt{\pi}$ for square of $\pi$ area. The length of the grain with $f_{u+1}$ dipoles is the module of the vector $(f_{u},f_{u+1})$ which is $L=\sqrt{\pi}(f_{u}^2+f_{u+1}^2)^{1/2}$ or simply $(f_{2u+1}\pi)^{1/2}$ . Folding  the strip into a ring, this length is the perimeter of a circle, then it is possible to estimate the number of points of the phyllotactic pattern in the domain enclosed by this circle.

Dipoles which are oriented along parastichies (refereed by $u$ in table~\ref{tab1})  are close to make the same angle with the radial direction (in absolute value as their orientations alternate).
Considering the description of grain boundaries  by a strip in a square lattice (figure \ref{fA1}), dipoles are represented along square edges and make a constant angle with the large strip side related to the slope of the strip $f_u/f_{u-1}$. The angle with the normal to the strip is $\textrm{arccot}(f_u/f_{u-1})\simeq \textrm{arccot}(\tau)\simeq 0.5535 \textrm{~rd} $.  This  is also the angle of a dipole with the radial direction for a strip refolded into the ring of a grain boundary if we consider that resulting distortions are small. In fact this is exactly the angle  with the radial direction done by the parastichies $u$ through the medium point of the grain boundary, whatever is the curvature.

There are  two correlated properties of inflation-deflation symmetry related to grain boundaries. The first one is the  organization of dipole on a grain boundary which is like an approximant. The other aspect, developed in \cite{sadocriviercharvolin} relates a grain boundary to the next one. There is a symmetry mixing spiral symmetry and inflation-deflation symmetry associating the two grain. Nevertheless such symmetry needs the scale invariance of the plane  and cannot be considered in curved geometries. In plane geometry the ratio of the radius of two successive grain boundaries is $\sqrt{f_{2u+3}/f_{2u+1}}$ converging toward $\tau$ for large Fibonacci numbers. This is a consequence of the inflation-deflation symmetry in plane geometry.
The structure of a grain boundary is only related to its rank $u$ through  Fibonacci number $f_u$ and does not depend on the space curvature. It is the same for spherical, plane and hyperbolic phyllotaxis.

\section{Appendix: The Poincar\'e disc representation of hyperbolic plane}
The Poincar\'e disc model, also known as the conformal disc model is a simple way to represent the hyperbolic plane.  It is
a mapping of the whole hyperbolic plane on the interior of a circle, so this limit circle represent infinity. It is a
conformal mapping respecting angles between geodesic lines which are represented by  arcs of circles that are orthogonal
to the limit circle~\cite{hilbert,thurston}.

A method to introduce the hyperbolic plane and its representation, is to refer to spherical geometry changing the
Gaussian curvature $\kappa=1/R^2$ into $-1/R^2$, where $R$ is the radius of curvature of the surface. By analogy with the
equation of a sphere embedded in the Euclidean 3D space, it would be tempting to write $ x^2+y^2+z^2=-R^2$ with an imaginary radius, but  the hyperbolic plane can not be entirely embedded in $\mathbb{R}^3$. So we need a representation of the hyperbolic plane
by a surface with a given metric. A good example is given by the Minkowski plane. Consider a 3D vector
$\textbf{y}=(y_0,y_1,y_2)$ in a 3D space with a metric such that the squared modulus of the vector is $\mid \textbf{y}
\mid^2=-y_0^2+y_1^2+y_2^2$. The equation $-y_0^2+y_1^2+y_2^2=-R^2$ defines a two sheets hyperboloid surface. A single
sheet, with the appropriate metric, is a model of a surface, with Minkowski metric, whose Gaussian curvature is $-1/R^2$. Geodesics of this surface are intersection  lines  between the hyperboloid and planes through the origin.

Using polar coordinates to describe the $ \textbf{y}$ vector we have $ y_0= R\cosh \varphi, y_1=R \sinh \varphi
\cos\vartheta, y_2=R \sinh \varphi \sin\vartheta$ for points on a sheet. The Minkowski metric $d\sigma^2=-dy_0^2+dy_1^2+dy_2^2$ is
with polar coordinates $d\sigma^2=R^2(d\varphi^2+\sinh^2 \varphi  d\vartheta^2)$. By comparison with spherical geometry
it is interesting to note that if we change  variables into complex variables $(y_0= \eta_0,y_1=i\eta_1,y_2=i \eta_2)$
and $\varphi=i \phi$ the metric look like the usual metric of a sphere $d\sigma^2=R^2(d\phi^2+\sin^2 \phi
d\theta^2)$. In spherical geometry, the parameter $R$ is the radius of the sphere. In hyperbolic geometry $R$ can not be consider as a radius of curvature, it is only the Gaussian curvature $\kappa=-1/R^2$ which is an intrinsical property of the surface.

The Poincar\'e disc representation is obtained by stereographic projection of one hyperboloid sheet unto the plane
$(y_1,y_2)$ with a projection pole $(y_0=-1,y_1=0,y_2=0)$.
We consider polar coordinates $(r,\vartheta)$ in the plane $(y_1,y_2)$
(that is in the Poincar\'e disc). The stereographic projection give $r=R \sinh \varphi/(1+y_0)$ or
$r=R\tanh(\varphi/2)$. Using hyperbolic trigonometric identities the metric is
$d\sigma^2=4R(dr^2+r^2 d\vartheta^2)/(1-r^2)^2$. This metric has the form of an Euclidean metric in polar coordinates divide by a function of $r$ only: it is locally an Euclidean metric, a proof that the representation is conform.

To obtain the equation of the spiral  allowing to build a phyllotaxis on the hyperbolic plane,
the calculation of the area enclosed in a circle of a given radius $R\varphi_\rho$ is needed.
This area is obtained using this metric. The perimeter of the circle   centered at the origin is $2\pi R \sinh \varphi$ and so the area enclosed by this circle
is $\int_0^{\varphi_\rho}2 \pi R^2 \sinh\varphi d\varphi$ which is:
\begin{equation}
2 \pi R^2(\cosh\varphi_\rho-1). \label{equaB1}
\end{equation}

\section{Appendix: A plane representation of the spherical phyllotaxis}

\begin{figure}[tpb]
\caption{Stereographic projection of a spherical phyllotaxis (part). This is given using equation (C1) with 2000 represented points and
$a=1/40$ so the total number of points on the sphere is $6400$. By effect of the projection size of cells seem to increase going outward, diverging for $n=6400$.}
\includegraphics{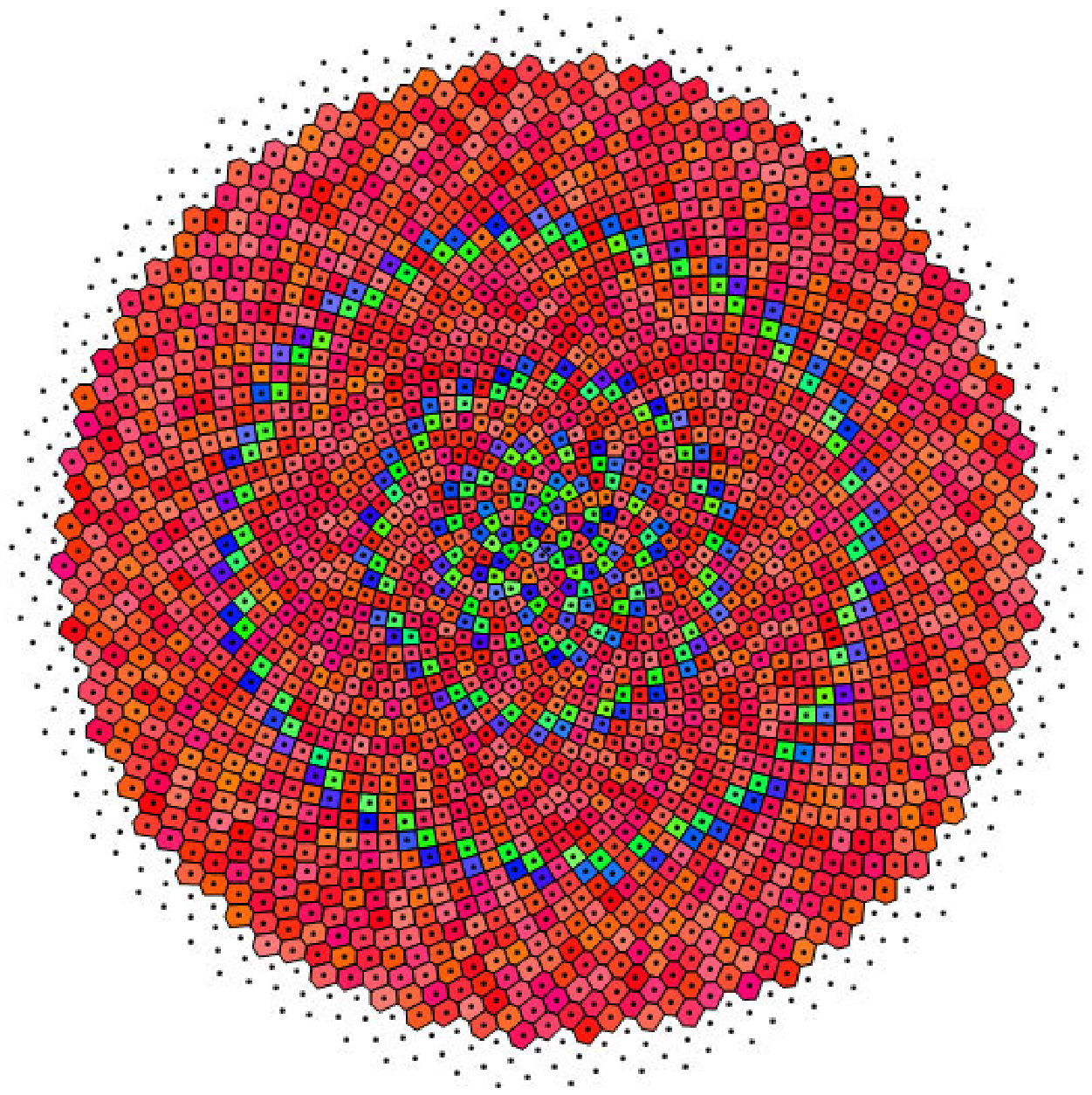}

\label{fC}
\end{figure}

 There is an other way
to describe the spherical phyllotaxis by analogy to the hyperbolic one, simply changing hyperbolic trigonometry into
spherical trigonometry. This give a representation on the plane of a spherical phyllotaxis, with an adjusted metric as shown in figure~\ref{fC}. In
fact it is a stereographic projection of the tiled sphere onto a tangent plane through north pole from a projection point on  the south pole.

The radial position for a point is given on the plane by
\begin{eqnarray}
r(s)=\tan \left(\frac{1}{2} \cos ^{-1}\left(1-a^2 s/2\right)\right). \label{equaC1}
\end{eqnarray}
This radial value is the radial coordinate on the plane and  the true spherical distance from the origin to the point depends on the spherical metric:
\begin{eqnarray}
d\sigma^2=4(dr^2+ r^2d\theta^2)/(1+r^2)^2. \label{equaC2}
\end{eqnarray}
 This is  a metric which has the form of an Euclidean metric in polar coordinates divided by a function  of $r$ only. This is practically useful to calculate distances: they are calculated as Euclidean distances on the representation, then corrected using the $r$ function. This is a good approximation if distances remain small compare to the radius of curvature. This locally Euclidean metric is related to the fact that the stereographic projection on the plane of the spherical tiling is conform.
The azimuthal position is the same as in the plane case: $2\pi s \lambda$ with $\lambda=1/\tau$.

Comparing the  Euclidean plane case to the plane representation of curved spaces like the Poincar\'e disc in the hyperbolic case or the stereographic map of the sphere, the main difference is the behaviour of the radial part $r(s)$ of the spiral equation. Evidentally if we consider the usual metric of the plane, and not the given metric of representations, the size of cells increases in the stereographic map and decreases on the Poincar\'e disc. It seems possible to generalize this observation to phyllotaxis where the size of cells is not constant. Any plane phyllotaxis build using a spiral equation in polar coordinates $(r(s),\theta(s))$ with $\theta(s)=2\pi s/\tau$ would have always the same grain boundaries. For instance this apply to composed flowers like sunflower where the size of florets increase from the core to the periphery due to growing, very similar to the plane representation of the spherical phyllotaxis but with the plane metric.

\end{document}